\documentclass[conference]{IEEEtran}
\IEEEoverridecommandlockouts

\makeatletter
\def\ps@IEEEtitlepagestyle{
        \def\@oddfoot{\mycopyrightnotice}
        \def\@evenfoot{}
}
\def\mycopyrightnotice{
        {\footnotesize
                \begin{minipage}{\textwidth}
                        \centering
                        \textcopyright~2024 IEEE.  Personal use of this material is
permitted.  Permission from IEEE must be obtained for all other uses, in
any current or future media, including reprinting/republishing this
material for advertising or promotional purposes, creating new
collective works, for resale or redistribution to servers or lists, or
reuse of any copyrighted component of this work in other works.
                \end{minipage}
        }
}

\usepackage[T1]{fontenc}
\usepackage{xspace}
\newcommand{\approach}{\emph{AdapTA}\xspace}

\newcommand{\fsr}{\textit{FSR}}
\newcommand{\dsr}{\textit{DSR}}

\usepackage{array}

\usepackage[normalem]{ulem} 
\usepackage[svgnames]{xcolor}
\usepackage{colortbl}  
\usepackage{hyperref}
\usepackage{comment}
\usepackage{amsmath,amssymb,amsfonts}
\usepackage{algorithmic}
\usepackage{graphicx}
\usepackage{textcomp}
\usepackage{xcolor}
\usepackage[numbers,sort&compress]{natbib} 
\usepackage{cleveref}
  
\usepackage{listings}
\usepackage{ifthen}
\usepackage{amssymb}
\usepackage{amsfonts}
\usepackage{etoolbox}
\robustify\cite

\newboolean{showcomments}
\setboolean{showcomments}{true} 
\ifthenelse{\boolean{showcomments}}
  {\newcommand{\nb}[2]{
    \noindent\fcolorbox{gray}{yellow}{\bfseries\sffamily\scriptsize#1}
    {\sf\small$\blacktriangleright$\textit{#2}$\blacktriangleleft$}
   }

   \newcommand{\ins}[1]{\textcolor{blue}{#1}} 
   \newcommand{\del}[1]{\textcolor{red}{\sout{#1}}} 
   \newenvironment{insb}{\par\color{blue}}{\par} 
  }
  {\newcommand{\nb}[2]{}
   
   \newcommand{\ins}[1]{#1} 
   \newcommand{\del}[1]{} 
  }

  \newcommand{\sam}[1]{\nb{Samira}{\footnotesize #1}}
\newcommand{\ric}[1]{\nb{Ricardo}{\footnotesize #1}}
\newcommand{\ant}[1]{\nb{Antonia}{\footnotesize #1}}

\def\BibTeX{{\rm B\kern-.05em{\sc i\kern-.025em b}\kern-.08em
    T\kern-.1667em\lower.7ex\hbox{E}\kern-.125emX}}
\begin{document}
\lstdefinelanguage{XML}
{
  basicstyle=\ttfamily\scriptsize,
  morestring=[b]",
  moredelim=[s][\bfseries\color{Maroon}]{<}{\ },
  moredelim=[s][\bfseries\color{Maroon}]{</}{>},
  moredelim=[l][\bfseries\color{Maroon}]{/>},
  moredelim=[l][\bfseries\color{Maroon}]{>},
  morecomment=[s]{<?}{?>},
  morecomment=[s]{<!--}{-->},
  commentstyle=\color{DarkOliveGreen},
  stringstyle=\color{blue},
  identifierstyle=\color{red},
  numbers=none
}

\title{An Adaptive Testing Approach Based on Field Data\\
}

\author{
\IEEEauthorblockN{Samira Silva$^{1,3}$, Ricardo Caldas$^2$, Patrizio Pelliccione$^1$, and Antonia Bertolino$^{1,3}$}
\IEEEauthorblockA{$^1$ \textit{Gran Sasso Science Institute (GSSI)}, 
L'Aquila, Italy, 
name.surname@gssi.it\\
$^2$ \textit{Chalmers University of Technology}, 
Gothenburg, Sweden, 
name.surname@chalmers.se\\ 
$^3$ \textit{ISTI-CNR}, 
 Pisa, Italy, 
 name.surname@isti.cnr.it 
 }
 }

\maketitle

\begin{abstract}
The growing need to test systems post-release has led to extending testing activities into production environments, where uncertainty and dynamic conditions pose significant challenges. Field testing approaches, especially Self-Adaptive Testing in the Field (SATF), face hurdles like managing unpredictability, minimizing system overhead, and reducing human intervention, among others. Despite its importance, SATF remains underexplored in the literature. This work introduces AdapTA (Adaptive Testing Approach), a novel SATF strategy tailored for testing Body Sensor Networks (BSNs). BSNs are networks of wearable or implantable sensors designed to monitor physiological and environmental data. AdapTA employs an ex-vivo approach, using real-world data collected from the field to simulate patient behavior in in-house experiments. 
Field data are used to derive Discrete-Time Markov Chain (DTMC) models, which simulate patient profiles and generate test input data for the BSN. The BSN’s outputs are compared against a proposed oracle to evaluate test outcomes. AdapTA's adaptive logic continuously monitors the system under test and the simulated patient, triggering adaptations as needed. Results demonstrate that AdapTA achieves greater effectiveness compared to a non-adaptive version of the proposed approach across three adaptation scenarios, emphasizing the value of its adaptive logic.
\end{abstract}


\section{Introduction}\label{sec:intro
}

Software testing has traditionally been regarded as a process focused on identifying faults during the development phase~\cite{Avizienis2004}. As its activities are typically carried out within the development environment, this task is named in \cite{Bertolino2021} as in-house or in-vitro software testing. However, in recent years, there has been a growing awareness among academic researchers and industry professionals of the need to continue testing after the software has been released and is in use (e.g.,~\cite{Fredericks2014,Fitzgerald2017,Bertolino2012,Baresi2010}). As a matter of fact, a significant number of failures reported in production are linked to problems that would be hard, if not impossible, to identify through in-house testing \cite{Gazzola2017}, due to the vast complexity, continual evolution, and intricate interconnections of today's software-intensive systems.

The work in \cite{Bertolino2021}, which conducts a systematic review of the literature concerning field-based testing techniques, categorizes Testing in the Field into online testing, offline testing, and ex-vivo testing. According to them, test cases can be executed in various ways depending on their proximity to the production environment. When they run directly on the production system itself, it is referred to as online testing. Alternatively, if the testing occurs on a separate instance that operates alongside the production environment, it is referred to as offline testing. Lastly, ex-vivo testing involves executing test cases in-house but using real-world data collected from the field for a more authentic evaluation. Also, in their work, the authors noticed that many of the collected works adopt some adaptation strategy in order to deal with uncertainty or confront emergent behaviors of the System Under Test (SUT).

We adopt here the term "Self-Adaptive Testing in the Field", or SATF, for short, which was first proposed in \cite{silva2024self}. That work, based on a literature review in the field, proposes a taxonomy and a definition for SATF, as "any type of testing activities performed in the field, which have the capability to self-adapt to the different needs and contexts that may arise at runtime"~\cite{silva2024self}. Field-based testing is crucial for identifying faults that may be overlooked during in-house testing \cite{Bertolino2021}. However, this becomes even more critical for systems that evolve over time, where traditional testing methods may not effectively capture emerging issues.
The work in \cite{silva2024self} also argues that the same factors driving the shift of testing activities from in-house to the field, such as addressing dynamism, context-dependence, and uncertainty, also support the need for self-adaptive testing approaches. That is to say that self-adaptability is a key characteristic in the execution of field testing activities.



An example of systems that evolve over time is the Body Sensor Networks (BSNs). A BSN consists of a network of wearable or implantable sensor nodes carefully positioned on or inside the human body to track physiological parameters or surrounding environmental factors~\cite{8360716}. Such systems are frequently employed in healthcare to monitor critical health metrics, including heart rate, body temperature, and glucose levels. Furthermore, BSNs find applications in areas such as the healthcare industry ~\cite{alrige2015toward,aziz2006pervasive,lee2015design}, the military-industrial sector~\cite{tatbul2004confidence}, sports and entertainment~\cite{conroy2009tennissense, pansiot2010swimming, burchfield2010framework}, and the social public field~\cite{osmani2007self, wai2010implementation}. BSNs are good exemplars of evolving systems since, for example, their sensors may be deactivated/activated over time. Their environment, which includes the monitored patient, may also change, for example by replacing a patient with a normal Body Mass Index (BMI) that was previously connected to the sensors with an Obesity-3 one. Finally, the health conditions of a patient also change over time, requesting different levels of attention.  


The review in \cite{silva2024self} concludes that SATF remains an immature topic, still attracting insufficient attention by software testing researchers, despite the wealth of open challenges. One emerging challenge is how to handle the uncertainty that is inherently encountered when testing in the field, where one is going to face many diverse operational scenarios that are difficult to predict.
In this work, we contribute to addressing this challenge by proposing a novel SATF approach, called AdapTA (Adaptive Testing Approach), to test BSNs. 

AdaptA is an ex-vivo testing approach, in which actual patient data, which we here call Field Sensor Reading (\fsr{}), is \textit{collected} from the field and \textit{preprocessed} to be transformed into Discrete-Time Markov Chains
(DTMCs) (\cite{trivedi2008probability}, Ch.7) 
that represent the condition of each sensor over time. Next, these DTMCs are used to \textit{simulate} patient profiles. For each patient profile, we run its DTMCs and input the obtained data, the DTMC Sensor Reading (\dsr{}), into the BSN, while a \textit{Monitor} component is responsible for monitoring the environment (e.g., the patient profile and condition) and the SUT (e.g., activation or deactivation of sensors). If a change occurs in either of them that triggers adaptation, the \textit{Analyse} component decides which adaptation should be made based on the changed object. Then, a \textit{Plan} for the adaptation of Test Cases, Oracle, or Test Strategy is made and \textit{Executed}. 
The \textit{Test Strategy} component decides when it is time to launch a test session, 
during which the \textit{BSN Outcomes} together with \dsr{}s are \textit{logged}, 
or otherwise the \textit{Patient Simulation} is continued.
When testing, based on each \textit{\dsr{}}, the \textit{Expected Outcome} is \textit{computed} and \textit{compared} to the actual \textit{BSN Outcome},
to decide whether the Test Cases (pairs of \dsr{} and \textit{expected outcome}) pass or fail.
For the empirical evaluation, we use \approach{} to test a self-adaptive system from the literature, the SA-BSN (Self-Adaptive Body Sensor Network)~\cite{gil2021body}, which already has been used in other works~\cite{rodrigues2018learning,silva4876283different}.

In summary, this work provides the following original contributions:
\begin{itemize}
    \item We introduce \approach, a novel self-adaptive ex-vivo approach to test BSNs;
    \item We describe three adaptation scenarios by providing the trigger activation and the adaptation policy for each of them; 
    \item We include extensive experimentation (covering the three scenarios, the baseline for each of them, and replicating each experiment 5 times);
    \item We perform a thorough statistical analysis of the results using both the Mann-Whitney U Test and the Vargha-Delaney A measure. 
\end{itemize}

The paper is organized as follows. 
Section \ref{sec:method} describes \approach{}; Section \ref{sec:exp:subj} outlines the system under test and the database with data from the field we employed; in Section~\ref{sec:study} we discuss the study setting and in Section~\ref{sec:results} the results and threats to validity; finally, in Section~\ref{sec:relatedwork} we overview the related literature and in Section~\ref{sec:conc} we wrap up and hint at promising future research directions.

\section{Adaptive Testing Approach: \approach{}}\label{sec:method}

\begin{figure*}[ht]
    \centering
    \includegraphics[width=0.95\textwidth]{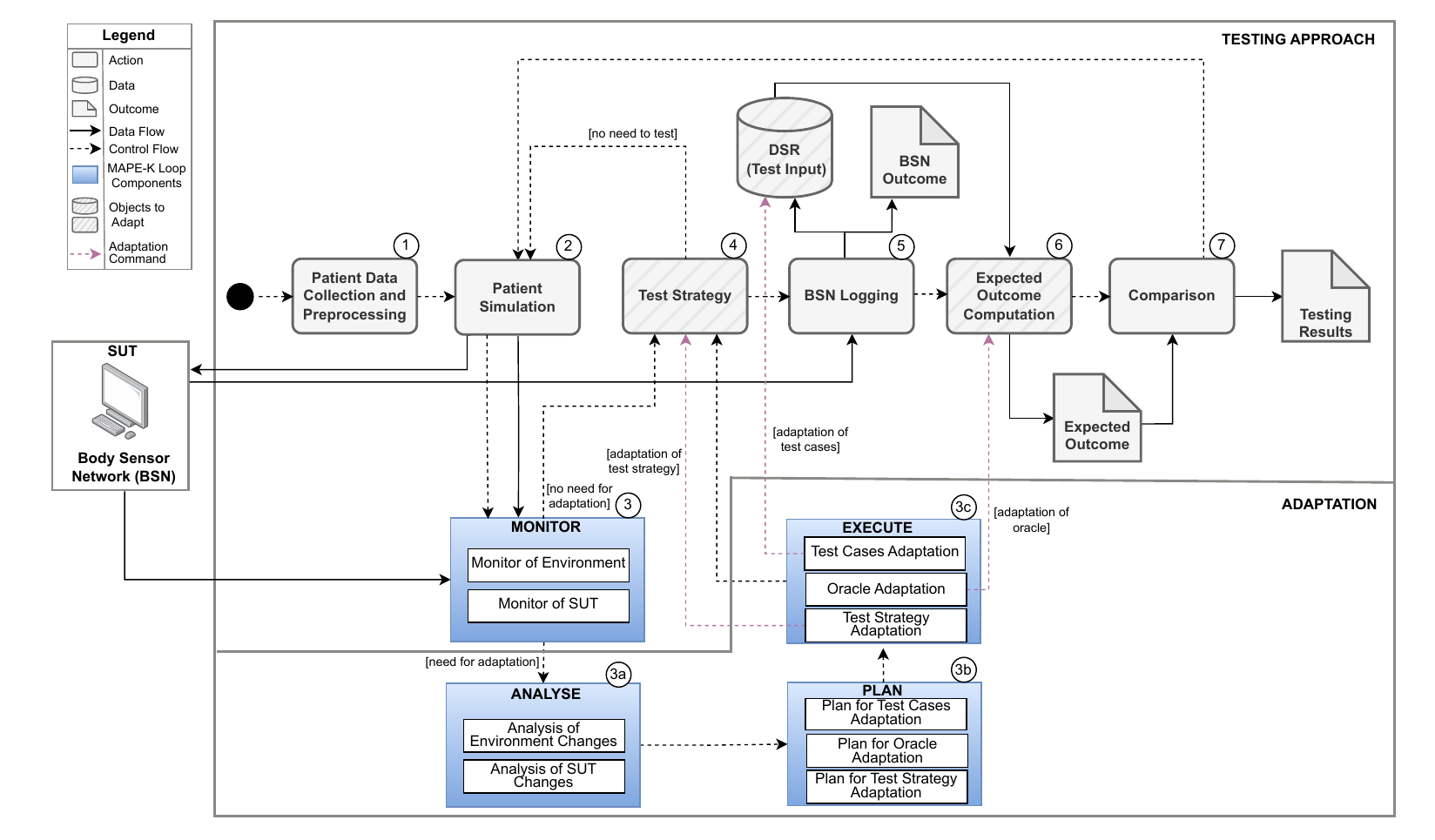}    
  
    \caption{
    Adaptive Testing Approach (AdapTA).} 

    
    \label{fig:method}
\end{figure*}

In this section, we introduce the Adaptive Testing Approach, or \approach for short. For better clarity, we first describe the basic Testing Approach without adaptation, and then we present the components of the Adaptation logic. 
For illustration, we refer to Figure \ref{fig:method} in which the stages of \approach are connected through either dashed arrows that indicate the flow of stages or commands (for commands, we use specific guards), or continuous arrows that indicate the transmission of data. 
Finally, we also present the three adaptation scenarios handled in~\approach{}. 

\subsection{Testing Approach}
In this section, we describe the seven basic components of the testing approach. The sole use of them, without the adaptation components, makes up the non-adaptive version of \approach{}, which will later serve as a baseline.

\subsubsection{Patient Data Collection and Preprocessing}
With reference to Figure \ref{fig:method}, Stage \textcircled{\small{1}} consists of collecting and preprocessing vital signs data from patients that will be later used to generate the test inputs to the BSN. Note that this is a stage that we do only once, as shown in the figure (i.e., it is outside of the iteration).
As said,  we here perform ex-vivo testing~\cite{Bertolino2021}, which means that we get actual data from the field (\fsr{}s), but use this data to generate DTMCs representing a patient, and later run the DTMCs to generate \dsr{}s and perform test sessions in-house. 
As an aside comment, online testing would correspond to directly launching the test cases while the BSN is in use and monitors a patient.
However, due to the safety-critical nature of BSNs, it would be of course risky to conduct testing directly online. 

Field data could be either collected directly from the patients and then promptly processed for testing in the subsequent steps, or otherwise they could be collected over some period of time, and saved into a database for later processing. 
This work applies the second case (data are taken from a database), even though either case would work for \approach{}.

An advantage of deferring the testing sessions after collecting a bulk of data into a database is that we can merge data coming from different patients so that we can obtain more general \textit{patient profiles}. One profile corresponds to a category of patients yielding similar physical and health-related characteristics (e.g., age, gender, height, 
weight, or some monitored pathology). We can thus test the behavior of the BSN over different patient profiles; in traditional functional testing, this roughly corresponds to testing over different equivalence classes (e.g.,~\cite{jorgensen2013software}, Ch.6). Therefore, we could test for instance the behavior of the BSN when monitoring a young healthy athlete or when instead monitoring an elderly patient with critical pathologies.
Patient profiles are modeled
through Discrete-Time Markov Chains (DTMCs).  Precisely, for a considered patient profile, a DTMC is constructed for each vital signal (i.e., each sensor of the BSN). The DTMCs are designed to represent the possible states of a sensor (defined by the selected values) and the transition probabilities which
reflect the likelihood of changes in vital signals over time. For example, the DTMC for body temperature would model the transitions from a low-risk state (e.g., [36$^{\circ}$C, 38$^{\circ}$C]) to a medium-risk state (e.g., [38$^{\circ}$C, 41$^{\circ}$C]), among others. This approach allows for capturing variations across differing profiles, such as the higher probability of an elderly patient progressing to a high-risk scenario compared to a youthful patient, or the increased risk associated with obesity compared to normal weight.  

\subsubsection{Patient Simulation} Stage \textcircled{\small{2}}, Patient Simulation, consists in, for each patient profile, running the DTMCs of its sensors to simulate a patient. In \approach{}, a patient is simulated during a certain amount of time $m$, and the data produced, \dsr{}, is sent as input to the BSN and the Monitor component. 

\subsubsection{Monitor}
The monitoring component, in Stage \textcircled{\small{3}}, must efficiently collect and check data from both the BSN and Patient Simulation (environment).
The Monitor is also a component of the MAPE-K loop, as explained later in Section~\ref{sec:adapt}. In case it observes changes in the SUT (the BSN) or in the environment (the Patient), we proceed to the next stage of the loop (Analyse - Stage \textcircled{\small{3a}}). Otherwise, if no adaptation is needed, it directs the flow to Test Strategy (Stage~\textcircled{\small{4}}). For the baseline, the latter is always chosen, as explained later in this work.

\subsubsection{Test Strategy}\label{subsec:ts}
The Test Strategy (Stage \textcircled{\small{4}} in Figure~\ref{fig:method}) decides when to test based on some condition. If this condition is satisfied, we move to the next stage, BSN Logging \textcircled{\small{5}}, and start testing. Otherwise, we go back to Stage~\textcircled{\small{2}} 
{(``No need to test'')} 
and continue simulating the patient. Testing may be triggered periodically, that is, testing is launched with a determined frequency and duration, or triggered by some event (e.g., SUT adaptation). We then decided to use a periodical trigger as a default strategy which means that we test for $t$ seconds every $w$ seconds. Thus, the condition to be satisfied here to test is being within the testing interval. 

\subsubsection{BSN Logging}
Stage \textcircled{\small{5}} involves recording data from the BSN during its execution. 
This process allows the data used by the BSN, i.e., DTMC sensor readings (\dsr{)}, and its corresponding output provided by the BSN, to be stored for later use in testing. The BSN outcome may consist, for example, of the overall risk level of a patient according to the BSN at a determined instant of time.

\subsubsection{Expected Outcome Computation}

In Stage \textcircled{\small{6}}, the Expected Outcome, referred to as the Oracle, is derived from the DTMC sensor readings (\dsr{}s). 
Specifically, if the BSN outcome is, for example, the overall risk level of a patient at a given moment, this stage comprises predicting it based on the data provided in the \dsr{}. The pair of \dsr{} and its Expected Outcome make up a Test Case.

\subsubsection{Comparison}
Finally, in Stage \textcircled{\small{7}}, the BSN Outcome is compared with the Expected Outcome (i.e., the Oracle) for a specific \dsr{}. In this stage, a comparison function that computes the distance between the BSN and Expected outcomes may be defined. 
By comparing them, we are able to know whether a determined Test Case passes or not. \approach{} runs for a predefined duration, denoted as $m$, which corresponds to the duration of the patient simulation, as previously anticipated.

\subsection{Adaptation}\label{sec:adapt}

The adaptation logic of \approach{}, stages \textcircled{\small{3a}}, \textcircled{\small{3b}}, and \textcircled{\small{3c}}, is inspired by the reference architecture for SATF approaches proposed in \cite{silva2024self}.
Thus, our adaptation logic also uses the MAPE-K loop, the most influential reference control model for autonomic
and self-adaptive systems \cite{Kephart2003}. The Knowledge components are the \dsr{}, and the BSN and Expected Outcomes. The remaining components are described as follows. As will be described in more detail in the following, the entire MAPE-K loop is executed only when the monitor identifies the need for adaptation. Besides, even if the Monitor is part of the MAPE-K loop, we here make it part of the testing approach since it is needed even if the adaptation logic is removed from \approach{}.

\subsubsection{Analyse}
The Analysis component is responsible for deciding which adaptation should be planned based on the monitored data. This data may be environmental (patient) or SUT (BSN) changes. The Analysis component then informs the Plan component about its decision. 

\subsubsection{Plan}
Knowing which adaptations should be made, the Plan stage formulates a plan with appropriate and concrete actions that make up these adaptations. For \approach, the possible adaptation mechanisms are Test Cases Adaptation, Oracle Adaptation, and Test Strategy Adaptation. 

\subsubsection{Execute}
The Execute stage consists of performing the adaptation activities described in the plan. For \approach{}, we consider three possible objects to adapt: Test Strategy, Test Cases (more specifically, the DSRs), and the Oracle (i.e., the function that computes the expected outcome). The Execute component launches an adaptation function to change these objects following the plan. This adaptation function may change DSR values or replace the Testing Strategy or the Expected Outcome Computation functions with a new function according to what has been planned.


\subsection{Scenarios}\label{sec:method:scen}
In this work, we propose three adaptation scenarios to illustrate the use of \approach{}. Table \ref{tab:scenarios} summarizes them by providing the adaptation trigger, a more concrete description for it, the adaptation policy, and the object to adapt. We describe them in detail as follows. 

    \begin{table*}[!ht]
\centering
{
\caption{\label{tab:scenarios}Adaptation Scenarios for AdapTA.}
\begin{tabular}{l|ll|ll}
\textbf{Scenario} & \textbf{Adaptation Trigger}  & \textbf{Trigger Description}      & \textbf{Adaptation Policy}           & \textbf{Object to Adapt} \\ \hline


S1                & Change in the SUT         & Activation/Deactivation of sensors             & Adjust the test cases & Test Cases         \\
S2                & Change in the environment & Change in the patient profile          & Adjust the oracle                    & Oracle                        \\ 
S3                & Change in the environment & The patient is in very critical health condition    & Adjust the frequency of testing                   & Test Strategy                        \\

\end{tabular}
}
\end{table*}

\subsubsection{Scenario 1 (S1) - Activation/Deactivation of Sensors}

\begin{itemize}
    \item \textbf{Description:} Let us suppose that the battery for a determined sensor is too low. Then, the BSN adapts itself by turning off this sensor. If we are testing this system, the test cases from this moment on should not include this sensor anymore. Also, the oracle, when computing the expected outcome, should not consider it. The opposite applies when a sensor gets activated. 
    \item \textbf{Adaptation Policy:} At runtime, if a sensor is deactivated, the test cases to be exercised should not consider it. Also, the oracle will not use it to compute the expected outcome. 
\end{itemize}
\subsubsection{Scenario 2 (S2)- Different Patient Profiles}

\begin{itemize}
    \item \textbf{Description:} Suppose that the sensors have been detached from a determined patient and attached to a different one with different health conditions (i.e., patient profile). The oracle should adapt to handle this change and consider the characteristics of the new patient to compute the expected outcome.  
    
    \item \textbf{Adaptation Policy:} If sensors are attached to a new patient, the oracle should take into consideration their health conditions (i.e., patient profile) to determine the expected outcomes for their sensor readings.
    
\end{itemize}

\subsubsection{Scenario 3 (S3) - Patient in Critical Condition}

\begin{itemize}
    \item \textbf{Description:} Testing the BSN when the patient is in a very critical condition is needed first for patient safety reasons since the BSN must work properly in this scenario. Also, this is a scenario in which an alarm is launched, 
     and a doctor/nurse should be called. Therefore, we do not want to require human resources, which are normally something very scarce in hospitals (and have a high cost) unnecessarily. This adaptation scenario allows an on-demand test strategy and it is useful when resources are limited. 
    
    \item \textbf{Adaptation Policy:} Whenever at least two of the sensors monitoring a patient report ``high risk", testing is triggered. 
 
\end{itemize}

\section{Experiment Subjects}\label{sec:exp:subj}

In this section, we describe the subjects involved in our experiments, including the system under test, the SA-BSN, and the MIMIC II Clinical Database, the database containing data from the field employed.

\subsection{Self-Adaptive Body Sensor Network (SA-BSN)}\label{sabsn}

In our experiments, we test the publicly available Self-Adaptive Body Sensor Network (SA-BSN)~\cite{gil2021body}\footnote{\url{https://github.com/rdinizcal/sa-bsn}}, which monitors a patient's health and detects emergencies in real time. The system includes six sensors: ECG for heartbeat and electrocardiogram, pulse oximeter (Oxi) for blood oxygen saturation, thermometer (Term) for body temperature, sphygmomanometer for blood pressure (Abpd and Abps), and a glucose sensor (Glc) for blood glucose levels. Table \ref{tab:ranges} shows the defined risk level ranges for these sensors.

SA-BSN is self-adaptive, adjusting the central hub (i.e., a component to fuse vital signs and classify the overall health situation of the
patient), sensor frequency, or activating/deactivating sensors based on environmental and internal changes. We used a version of SA-BSN that we fixed and made available in our replication package\footnote{\url{https://github.com/samirasilva/AdapTA_Paper_AST}}. The original version had three bugs affecting sensor reading printing functions, the DTMC state transition function, and the lack of message sending to other ROS (Robot Operating System) nodes whenever a sensor is activated/deactivated. These issues, which impacted result accuracy, were reported to the developers and resolved in the version used in our experiments.
\begin{table}[!htb]
\caption{Data ranges and risk levels for the sensors of SA-BSN \cite{solano2019taming}}
\label{tab:ranges}
{\scriptsize
\centering
\resizebox{\columnwidth}{!}{%
\begin{tabular}{p{3.7cm}|p{4.7cm}}
\textbf{Sensor} & \textbf{Data Ranges}                            \\ \hline
Oxygen Saturation (Oxi)   & 100$>$low$>$65$>$medium$>$55$>$high$>$0                     \\ \hline
Heart Beat Rate (Ecg)    & 300$>$high$>$115$>$medium$>$97$>$low$>$ 85$>$ medium$>$70$>$high$>$0  \\  \hline
Temperature (Term)          & 50$>$high$>$41$>$medium$>$38$>$low$> $36$>$ medium$>$32$>$high$>$0    \\  \hline
Systolic Body Pressure  (Abps)  & 300$>$high$>$140$>$medium$>$120$>$low$>$0                   \\  \hline
Diastolic Body Pressure (Abpd)   & 300$>$high$>$90$>$medium$>$80$>$low$>$0                     \\  \hline
Glucose (Glc)          & 200$>$high$>$120$>$medium$>$96$>$low$>$ 55$>$ medium $>$40$>$high$>$20 \\  
\end{tabular}
}}
\end{table}

\subsection{MIMIC II Clinical Database}

In this work, we use data extracted from version 2.6 of the MIMIC II Clinical Database~\cite{saeed2011multiparameter}, specifically the subset used in the PhysioNet/Computers in Cardiology Challenge. The dataset includes records of 12,000 patients aged 16 or older, all of whom had an initial Intensive Care Unit (ICU) stay of at least 48 hours. The data contains admission information such as age, gender, height, ICU type, and initial weight, along with up to 41 variables recorded during the first 48 hours after ICU admission. However, not all variables are available in every record. The physiological measurements we use for this work include arterial blood pressure (systolic and diastolic), heart rate, oxygen saturation, and temperature. These variables are essential for monitoring and assessing the condition of ICU patients.
\vspace{-0.4cm}
\section{Experiment Setting}\label{sec:study}
In this section, we discuss the 
setting of the experiments we conduct to 
evaluate the effectiveness of \approach{} in failure detection across the three proposed scenarios. 
Specifically, we measure the Passing Test Case Rate (PTCR) for the three scenarios of \approach{} and their corresponding baselines.
PTCR consists of the percentage of test cases that pass in the SUT. Therefore, the lower the PTCR, the better, indicating greater effectiveness in detecting failures. The experimental procedure was designed following the guidelines outlined in~\cite{wohlin2012experimentation} and adheres to the seven stages detailed in Section~\ref{sec:method}.

In the remainder of this section, we describe the implementation of the Testing Approach components and the proposed scenarios. Details about the implementation of the adaptation components are embedded in Section~\ref{sec:study:scen}. Finally, we also describe the baseline for each scenario and the experimental setup. 

\subsection{Testing Approach}
\subsubsection{Patient Data Collection and Preprocessing}

To derive the patient profiles corresponding to the equivalence classes on which we test the SA-BSN, we first selected the attributes according to which the \fsr{}s could be grouped. Based on our study of the entries recorded in the MIMIC II database, 
we selected
two attributes relative to the patients, namely, Age and Body Mass Index (BMI). 
In the literature, the categorization of patient profiles by age is not defined by a unique standard but by different standards depending on the application. For the sake of this study, we defined our own categorization as follows:
Youth (16y-29y), Adult (30y-59y), and Elderly (60y+).
The BMI is defined by:

\begin{equation}
    BMI = \frac{weight(kg)}{height(m)^2}
\end{equation}

It yields a well-known categorization that was first proposed in \cite{quetelet1842treatise}, i.e.,: Underweight (0-18.4), Normal weight (18.5-24.9), Overweight (25-29.9), Obesity 1 (30-34.9), Obesity 2 (35-39.9), and Obesity 3 (40+). 
In addition, we use a third categorization according to the Intensive Care Unit (ICU) attributes of the records, as clinical factors influencing the patients' vital signals. The ICU categories, as provided in the database,  are: Cardiac Surgery Unit, Coronary Care Unit, Medical ICU, and Surgical ICU.

{After categorizing the field data according to the patient profiles, we generate a DTMC model for each sensor belonging to a patient profile. The DTMC states are the ranges that represent the risk levels for the sensor considered. The transitions in the DTMCs correspond to the probability of changing from one risk level to another. The transformation of the field data into DTMCs is done by: (1) using as states the same risk level ranges employed by our SUT, the SA-BSN (as defined in~\cite{solano2019taming}), and (2) assigning transition probabilities in the DTMCs based on the frequency with which these transitions occur in the field. The likelihood of a signal moving from one state to another may vary depending on the considered profile. For example, for the Temperature sensor, the transitions going to the low-risk state, which is fixed to [36$^{\circ}$C, 38$^{\circ}$C], may present a high probability in the DTMC for healthy patients.}

It is important to highlight that, in general, it is challenging to attribute semantics to the states as we do in this explanation. For example, knowing that [38.1$^{\circ}$C, 41$^{\circ}$C] represents a medium risk to elderly patients requires domain expertise.

The transition probabilities must be computed for each sensor of each patient profile. To calculate these probabilities, for a determined sensor and patient profile, we compute a weight that is the number of transitions from one state to the others in the field data. 
Then, we normalize the weights of the transitions in each DTMC by dividing each weight by the maximum weight.

\subsubsection{Patient Simulation} The patient simulation consists in running {the DTMCs for a patient profile}
for $m$ seconds. Due to limited resources and time constraints, we restrict $m$ to 3600 seconds. {The DTMCs of a patient profile execute to} produce input for the BSN, the DTMC Sensor Readings (\dsr{}s).

\subsubsection{Monitor} The monitor is responsible for monitoring which is the current patient profile, the DSRs coming from the Patient Simulation, and the SA-BSN. Then, it decides whether there is a need for adaptation based on these observed objects. If the current patient profile or the patient conditions are critical, it activates the \textit{analysis} of environmental changes. On the other hand, if it observes that the SA-BSN is adapting to deactivate/activate sensors, it activates the \textit{analysis} of SUT changes is performed. Otherwise, if there is no need for adaptation, it directs the flow to Test Strategy.

\subsubsection{Test Strategy}
Test Strategy is responsible for determining when it is time to test. As a default strategy, testing runs for $t=60$ seconds, pauses for $w=300$ seconds, and then restarts in a continuous cycle. {Therefore, when reaching the Test Strategy stage, if it is testing time, we proceed to the next stage (BSN Logging). Otherwise, if it is break time, we return to the Patient Simulation stage.}
 
 \subsubsection{BSN Logging}
During a testing session, we collect and store all \dsr{}s that were input into the SA-BSN. Additionally, we capture the BSN Outcome, representing the patient’s risk level determined by the SA-BSN based on the \dsr{} values. For the SA-BSN, the possible patient's risk levels are: "Very Low Risk," "Low Risk", "Moderate Risk", "Critical Risk", and "Very Critical Risk". 

\subsubsection{Expected Outcome Computation}
The Expected Outcome Computation consists in determining the Expected Outcome (oracle) for each \dsr{} logged. As a default oracle, we use the oracle defined by our previous work in \cite{silva4876283different} for testing the SA-BSN. Table~\ref{tab:oracle} outlines this oracle.
\begin{table}[h]
\centering
\caption{Default Oracle Definition.}
\label{tab:oracle}
 \resizebox{\columnwidth}{!}{
\begin{tabular}{c|l|l}
\textbf{ID}&\textbf{Patient Risk Level} & \textbf{Condition} \\ \hline
1&Very Low Risk & All sensors at “Low Risk” \\ \hline
2&Low Risk & One sensor at “Medium Risk”; others “Low Risk” \\ \hline
3&Medium Risk & Two or more sensors at “Medium Risk”; none “High Risk” \\ \hline
4&Critical Risk & Exactly one sensor at “High Risk” \\ \hline
5&Very Critical Risk & More than one sensor at “High Risk” \\
\end{tabular}
}
\end{table}
Note that, to determine the oracle, we convert \dsr{} values into their corresponding risk levels (i.e., low, medium, or high) based on the thresholds defined by the SA-BSN, as outlined in Table \ref{tab:ranges}. Then, we tally the number of sensors at each potential risk level, and apply the rules defined in Table \ref{tab:oracle}. We chose to do this conversion because it simplifies the oracle and aligns more closely with the operation of the SA-BSN. However, \approach{} is flexible and its oracle may be easily adjusted for other BSNs that use different sensor types or ranges, or lack specific ranges altogether. We recall that each \dsr{} and its corresponding Expected Outcome make a Test Case. 

\subsubsection{Comparison}
The comparison consists in contrasting the BSN Outcome and the Expected Outcome (oracle) for a determined test case in order to decide if it passes or fails. This comparison is made by assigning a numerical ID for each of the possible Patient Risk Levels as shown in Table\ref{tab:oracle} and computing the absolute difference between them. A test case passes if the absolute difference between the IDs of the Expected and the BSN Outcomes is lower than 2 and fails otherwise. 

\subsection{Scenarios}\label{sec:study:scen}
This section outlines the implementation of the three adaptation scenarios introduced earlier in Section \ref{sec:method:scen}. 

\subsubsection{Scenario S1 - Activation/Deactivation of Sensors}
One of the adaptations performed by the SA-BSN is the dynamic activation or deactivation of sensors based on their battery level. A deactivated sensor switches to `charge' mode and reactivates automatically once it reaches a determined battery level.

In the SA-BSN, whenever a sensor is deactivated, subsequent sensor readings retain the last recorded value for that sensor. However, we believe this is not an effective strategy. For instance, if the majority of sensors are deactivated and their last recorded values are all within the "Low risk" range, it could hinder the accurate detection of a critical condition in the patient. On the other hand, if they are all in the "High risk" range, it may cause unnecessary alarms. 

{To address this issue, whenever the Monitor observes that a sensor is deactivated, it triggers the need for adaptation and the Analyse component decides that a change in the \dsr{}s should be made once it is stored. Then, the Plan component prepares a plan for the adaptation relying on replacing the outdated values from a deactivated sensor in the \dsr{} with a "deactivated" label. This value will be later excluded from the computation of the Expected Outcome. The Execute component is responsible for launching a function that does it. }

To better visualize the effects of adapting the testing in this scenario and increase the frequency of sensor deactivations/activations throughout the SA-BSN execution, we adjusted a parameter controlling the battery recharge/discharge rate, changing it from its default value of 0.05 to 0.65.

\subsubsection{Scenario S2 - Different Patient Profiles}

People with different health conditions exhibit different vital signs and physiological responses~\cite{cooney2010elevated,song2023body,innocent2013correlation}. In this scenario, we take this fact into consideration and propose the adaptation of the Expected Outcome Computation function to handle patients with different profiles.

For the experiments, we chose to investigate the six patient profiles of BMI, namely Underweight, Normal Weight, Overweight, Obesity 1, Obesity 2, and Obesity 3, since the literature shows that BMI is very correlated to vitals. We here focus on two vital signs that are particularly affected by the BMI, body pressure~\cite{dua2014body}, and heartbeat rate~\cite{zalesin2008impact}. Both signs, heartbeat rate (Ecg) and body pressure (Abps, Abpd) are influenced by the BMI whether the weight is too high (e.g., \cite{zalesin2008impact,cooney2010elevated, song2023body,dua2014body}) or too low (e.g., \cite{dua2014body,kwon2021incidence,wu2021prevalence}). The idea in this scenario is the computation of the oracle by weighting the Ecg, Abps, and Abpd risk levels according to the patient profile, among the ones categorized by BMI. {Thus, in this scenario we replace the default oracle described in Table~\ref{tab:oracle} with a new one, aiming to simplify the adaptation. We describe the computation of this new oracle as follows.} 


Each sensor may provide values in a Low, Medium, or High-risk range. To make easier and more effective the application of weights, we define a scoring system for sensor risk levels. Thus, for a sensor $i$ the transformation of its risk level $RL_{i}$ into a score $S_{i}$ is done as follows: 
\begin{equation}
S_i =
\begin{cases} 
5 & \text{if $RL_{i}$ is "Low"} \\ 
20 & \text{if $RL_{i}$ is "Medium"} \\ 
100 & \text{if $RL_{i}$ is "Hard"}  
\end{cases}
\end{equation}

Then, for each sensor $i$, we compute its contribution to the overall score $Overall\_Score$ based on its weight $W_{{i}}$. The formula for the overall score is:
\begin{equation}
\text{$Overall\_Score$} = \frac{\sum_{i=1}^{n} (W_i \times S_i)}{n}
\end{equation}
where $S_{{i}}$ and $W_{{i}}$ are the score and the weight of sensor $i$, respectively, and $n$ is the number of active sensors. {The default value for $W_{{i}}$ is 1.}

Finally, based on the calculated overall score $Score$, the thresholds for the Expected Outcomes are:

{\small
\begin{itemize}
    \item \textbf{ Very Low Risk}: \( Overall\_Score < 8 \)
    \item \textbf{ Low Risk}: \( 8 \leq Overall\_Score < 11 \)
    \item \textbf{ Moderate Risk}: \( 11 \leq Overall\_Score \leq 20 \)
    \item \textbf{ Critical Risk}: \( 20 < Overall\_Score \leq 36 \)
    \item \textbf{ Very Critical Risk}: \( Overall\_Score > 36 \)
\end{itemize}
}

{Whenever the Monitor observes changes in the patient profile (e.g., a simulated normal-weight patient was replaced by an overweight one), it triggers the need for adaptation. The Analyse component decides that a change should be made to the function that computes the expected outcome (the oracle). Then, the Plan component prepares the adaptation based on the new patient profile. The plan consists of the rules for defining $W_{{i}}$ based on the patient profile observed.}



{In this plan, the value of $W_{i}$ is defined as follows:}
\begin{itemize}
    \item For non-priority sensors (i.e., Oxi, Term, and Glc):
\end{itemize}
\begin{equation}
W_{i} = 1;
\end{equation}
\begin{itemize}
    \item For priority sensors (i.e., Ecg, Abps, and Abpd):
\end{itemize}
{\small
\begin{equation}
W_i =
\begin{cases} 
1 & \text{if patient profile is "Normal Weight"} \\
1.75 & \text{if patient profile is "Overweight" or ``Underweight"} \\
1.85 & \text{if patient profile is "Obesity 1"} \\
1.90 & \text{if patient profile is "Obesity 2"} \\
2 & \text{if patient profile is "Obesity 3"} 
\end{cases}
\end{equation}
}


This approach allows us to prioritize specific sensors while still considering the overall input from all sensors, yielding a more nuanced and actionable criticality assessment.
{Finally, the Execute component is responsible for launching a command to send the suitable value of $W_{{i}}$ to the Expected Outcome Computation.} This value of $W_{{i}}$ will be used in the computation of the oracle until the patient profile changes again. 

\subsubsection{Scenario S3 - Patient in Critical Condition}

In this scenario, we suppose that human and physical resources are limited and we aim to test the SA-BSN only if the patient is in a very critical situation. {Thus, we consider that the default testing strategy, described in Section \ref{subsec:ts} is replaced with a function that looks at a boolean parameter $ts$ indicating when it is time to test. This parameter is always set to "false" unless an adaptation changes its value.}

We implemented this scenario by {using the Monitor component to observe \dsr{}s coming from the Patient Simulation. If the patient is in a very critical situation, that is, more than one sensor value presents high risk level in a \dsr{}, it triggers the need for adaptation. Then, the Analyse component decides that an adaptation should be made to the Testing Strategy component. Next, the Plan component describes how this adaptation should be made, that is, the $ts$ should be changed to "true". The Execute component launches a function to execute this adaptation to the Testing Strategy.} This \dsr{} will be then considered for testing.

\subsection{Baseline}
In our experiments, we consider one Baseline for each scenario (i.e., S1, S2, and S3). The baseline consists of the testing components of \approach{} with the same configuration as in the respective scenarios but without performing adaptations. In other words,
referring to Figure~\ref{fig:method} and specifically to the Monitor component, this means to always take the "no need for adaptation" arrow. The baseline is described in the following. 

\subsubsection{Baseline - S1} As previously mentioned, when a sensor is deactivated, the SA-BSN considers the last value of the sensor when computing the BSN outcome. 
Thus, the non-adaptive version of scenario S1, i.e., the baseline, consists of doing the same and using this value to compute the Expected Outcome instead of ignoring deactivated sensors.  

\subsubsection{Baseline - S2} 
Scenario S2 considers that the oracle for patients with different profiles should also be different. Thus, it acts by assigning weights ($W_i$) to each sensor based on the patient profiles. The baseline for this scenario corresponds to removing the weights from the oracle and considering that all patient profiles should be treated equally. In practice, it means assigning $W_i=1$ regardless of the sensor or patient profile. 

\subsubsection{Baseline - S3}
In scenario S3, the default test strategy condition described in Section \ref{subsec:ts} is replaced with a condition that corresponds to testing whenever a patient is in a critical situation. The baseline for this scenario is exactly the use of the default test strategy condition, that is, a testing session is launched every $w$ seconds and performed for $t$ seconds, consisting of a periodical test strategy. In our experiments, we use $t=60$ and $w=300$, since these values seemed to provide a feasible balance between testing and pause.

\subsection{Experimental Setup and Procedure}

The experiments were conducted on a computer running Ubuntu 20.04 LTS, equipped with 16GB of RAM and a processor with 4 cores, each running at a speed of 2.3 GHz. Additionally, the ROS Noetic\footnote{\url{http://wiki.ros.org/noetic}} framework for Ubuntu 20.04 was required to execute the SA-BSN. The installation of the SA-BSN was carried out by following the instructions provided here\footnote{\url{https://github.com/lesunb/bsn}}. For a more realistic BSN experience, we also disabled a parameter in the SA-BSN that causes a sensor's drained battery to recharge instantly. Since \approach{} and the baseline utilize DTMCs and involve the generation of random numbers, they exhibit non-deterministic behavior in each execution. To ensure the validity of the experiments, we conducted 5 runs for each scenario proposed for both \approach{} and the baseline. 

\section{Experimental Results}\label{sec:results}
This section covers the experimental data, the statistical analysis conducted, and the potential threats to the validity of the experiments. The experiments evaluated the Passing Test Case Rate (PCTR) of \approach{} across the three proposed scenarios (S1, S2, and S3). The outcomes of the experiments demonstrate that \approach{}, when compared to the baselines, offers an effective solution for testing Body Sensor Networks. Precisely:

\begin{itemize}
    \item \approach{} achieved an average PTCR of $69.44\% \pm 6.08$ for S1, while the experiment with the baseline for the same scenario resulted in an average PTCR of $88.67\% \pm 2.28$.
    \item \approach{} achieved an average PTCR of $72.49\% \pm 5.67$ for S2, while the experiment with the baseline for the same scenario resulted in an average PTCR of $89.57\% \pm 2.75$.
    \item \approach{} achieved an average PTCR of $61.44\% \pm 3.20$ for S3, while the experiment with the baseline for the same scenario resulted in an average PTCR of $88.05\% \pm 1.29$.
\end{itemize}

The full experimental results are shown in Tables~\ref{tab:results} and~\ref{tab:resultsprofiles}. For data analysis, we adhered to the guidelines outlined in~\cite{de2019evolution,arcuri2014hitchhiker}. To ensure reproducibility, all scripts and datasets have been made available\footnote{\url{https://github.com/samirasilva/AdapTA_Paper_AST}}.
\vspace{-0.4cm}

\begin{table}[htb]
\centering
\caption{\label{tab:results} Passing Test Case Rate (PTCR) for the different scenarios of \approach{} and the baselines.}
\begin{tabular}{p{1.5cm}||p{.55cm}|p{.55cm}|p{.55cm}|p{.55cm}|p{.55cm}|p{.55cm}|p{.4cm}}
 &Ex1&Ex2&Ex3& Ex4& Ex5&\textbf{Avg}&Std   \\\hline
\hline


Baseline - S1&89.46&84.17&90.46&89.75&89.53& \textbf{88.67} &2.28  \\\hline
AdapTA - S1 &73.3&68.12&69.35&59.12&77.30&\textbf{69.44}&6.08  \\
\hline \hline
Baseline - S2 &92.22 &90.72 &89.68 &84.33 & 90.92& \textbf{89.57} &2.75   \\\hline
AdapTA - S2 &72.16&82.57 &73.36 &66.13&68.23
&\textbf{72.49}&5.67 \\\hline \hline
Baseline - S3 &86.29 &89.77&88.83&86.85&88.50& \textbf{88.05} & 1.29  \\\hline
AdapTA - S3 &56.76&65.65&61.98 &63.78&59.04&\textbf{61.44}&3.20 \\\hline

\end{tabular}
\end{table}
\vspace{-0.6cm}
\subsection{Empirical Data }

The data for analysis is gathered by logging sensor readings, the BSN outcome, and calculating the expected outcome. For both the proposed approach and the baseline in each scenario, the difference between the BSN outcome and the expected outcome is computed for each test case.

Due to the different behaviors of the approaches and the non-determinism of the SA-BSN, the number of test cases varies significantly across the different scenarios. For both \approach{} and the baseline in scenario 1, each of the 5 experiments, consisted of 13 patient profiles that run for one hour each, resulting in approximately 340637.4 test cases in each experiment. Scenario 2 contains only the BMI patient profiles, resulting in 6 patient profiles that run for one hour in each of the 5 experiments, and an average of 204027 test cases per experiment for both \approach{} and baseline. Finally, scenario 3 consists of the 13 patient profiles that run for one hour in each of the 5 experiments, but results in different amount of test cases for \approach{} and the baseline. While the baseline logs approximately 368442.6 test cases, \approach in this scenario collects an average of 212031.8 test cases, since it focus on testing the SA-BSN only when the patient is in a critical situation and not periodically as the baseline. 

\begin{table*}[!ht]

    \centering

 \caption{\label{tab:resultsprofiles} Passing Test Case Rate according to the different patient profiles for the Baselines and the scenarios we propose for \approach{}. }
 
\begin{tabular}{cc||cc|cc||cc|cc||cc|cc}
\multicolumn{1}{c}{}              & \multicolumn{1}{c||}{} & \multicolumn{2}{c|}{\textbf{Baseline-S1}} & \multicolumn{2}{c||}{\textbf{S1}} & \multicolumn{2}{c|}{\textbf{Baseline-S2}} & \multicolumn{2}{c||}{\textbf{S2}} & \multicolumn{2}{c|}{\textbf{Baseline-S3}} & \multicolumn{2}{c}{\textbf{S3}} \\ \cline{3-14} 
\multicolumn{2}{c||}{\textbf{Patient Profile}}             & \textbf{Mean}    & \textbf{Std}    & \textbf{Mean}   & \textbf{Std}   & \textbf{Mean}    & \textbf{Std}    & \textbf{Mean}   & \textbf{Std}   & \textbf{Mean}    & \textbf{Std}    & \textbf{Mean}   & \textbf{Std}  \\ \hline
\multicolumn{1}{c|}{\textbf{ICU}} & MedicalICU            & 88.77            & 2.89            & 55.82           & 11.89          & -                & -               & -               & -              & 86.44                & 2.82               & 60.60            & 6.04          \\
\multicolumn{1}{c|}{\textbf{ICU}} & CoronaryCareUnit      & 91.90             & 2.10             & 72.74           & 11.58          & -                & -               & -               & -              & 90.96            & 3.10             & 66.68           & 1.71          \\
\multicolumn{1}{c|}{\textbf{ICU}} & SurgicalICU           & 84.65            & 3.82            & 60.47           & 6.34           & -                & -               & -               & -              & 81.58            & 6.07            & 65.00              & 5.97          \\
\multicolumn{1}{c|}{\textbf{ICU}} & CardiacSurgeryUnit    & 93.57            & 2.42            & 68.44           & 12.71          & -                & -               & -               & -              & 90.06            & 4.48            & 57.54           & 6.27          \\ \hline
\multicolumn{1}{c|}{\textbf{Age}} & 16-29                 & 91.79            & 2.15            & 62.47           & 9.64           & -                & -               & -               & -              & 88.11            & 2.60             & 70.39           & 8.52          \\
\multicolumn{1}{c|}{\textbf{Age}} & 30-59                 & 85.77            & 5.97            & 65.44           & 9.74           & -                & -               & -               & -              & 84.18            & 6.05            & 64.52           & 7.13          \\
\multicolumn{1}{c|}{\textbf{Age}} & 60+                   & 84.86            & 5.42            & 61.54           & 4.46           & -                & -               & -               & -              & 86.03            & 5.99            & 63.12           & 5.51          \\ \hline
\multicolumn{1}{c|}{\textbf{BMI}} & Obesity3              & 88.46            & 1.65            & 69.83           & 7.72           & 85.58            & 2.01            & 61.28           & 4.42           & 82.10             & 4.66            & 59.64           & 5.05          \\
\multicolumn{1}{c|}{\textbf{BMI}} & Obesity2              & 92.44            & 4.77            & 67.88           & 8.01           & 87.99            & 8.00               & 68.15           & 10.00             & 83.87            & 2.29            & 61.14           & 2.54          \\
\multicolumn{1}{c|}{\textbf{BMI}} & Obesity1              & 93.06            & 3.92            & 65.66           & 7.63           & 90.09            & 2.09            & 70.69           & 4.17           & 89.44            & 5.33            & 51.76           & 10.85         \\
\multicolumn{1}{c|}{\textbf{BMI}} & Overweight            & 91.54            & 5.17            & 63.79           & 14.03          & 88.52            & 3.70             & 65.45           & 6.93           & 89.97            & 2.31            & 58.81           & 5.57          \\
\multicolumn{1}{c|}{\textbf{BMI}} & Normalweight          & 89.09            & 4.76            & 69.52           & 12.33          & 86.14            & 6.30             & 86.14           & 6.30            & 85.21            & 6.52            & 58.27           & 6.82          \\
\multicolumn{1}{c|}{\textbf{BMI}} & Underweight           & 88.62            & 5.61            & 67.60            & 11.05          & 89.40             & 4.54            & 66.45           & 7.38           & 88.97            & 3.52            & 62.79           & 7.43         
\end{tabular}
   
\end{table*}

\subsection{Data Analysis and Discussion}

To select an appropriate approach for statistical analysis, we thoroughly assess the characteristics of our results~\cite{de2019evolution}. After confirming that our data does not follow a normal distribution, we conduct the Mann-Whitney U Test~\cite{mann1947test} to evaluate at a 5\% significance level the null hypothesis that there are no statistically significant differences in PTCR between \approach{} and the baselines.  
Comparing the baseline approach with the proposed method, the p-value obtained across all scenarios 
was 1.219e-2, with a U statistic of 25. This 
confirms a significant difference in effectiveness between the two approaches, at the 95\% confidence level. 

A significant Mann-Whitney U test indicates that one testing approach stochastically dominates the other, 
however it does not specify the precise  relationship between the pairs of techniques. 
We conducted pairwise comparisons (testing approach with each scenario against its respective baseline) using the Vargha-Delaney effect size (A12 measure) \cite{vargha2000critique} following the Mann-Whitney U test. For all the pairs, the value of $A_{12}$ measure was equal to 1, which represents the maximum effect size, indicating an extremely strong domination of the Baseline by \approach{}. 

Based on the statistical analysis of the results presented in Table~\ref{tab:results}, \approach{} proves to be the most effective method for detecting failures compared to its non-adaptive counterpart, the baseline, as evidenced by the PTCR values. In Scenario 1, the significant difference in terms of PTCR between the Baseline and \approach{} provides evidence that using the last value received from a sensor when it gets deactivated at runtime may not be the best solution, since these old values could distort the actual patient's condition. Thus, adapting the test cases to no longer include these values may be more beneficial, and can capture more failures. In Scenario 2, the notable difference in PTCR between the Baseline and \approach{} suggests that adapting the oracle to handle different patient profiles is important since a more precise and personalized oracle may result in finding more failures. Finally, Scenario 3, with the most significant difference in terms of PTCR between \approach{} and the Baseline, demonstrates the importance of testing when the patient is in a critical situation. This may be the context in which the SA-BSN should perform best, since its malfunction, in this case, could lead to serious harm to the patient or even death. Therefore, proper attention should be given to this scenario.\\
\indent Table \ref{tab:resultsprofiles} reports the PTCR by patient profile and scenario for both \approach{} and the baseline. Since the adaptation performed in Scenario 1, the deactivation/activation of sensors, is not directly impacted by the patient profile, as in the other scenarios, we could not see a clear trend in 
the difference in PTCR between the Baseline and \approach{}. The "MedicalICU", "16-29", and "Overweight" patient profiles present the highest difference for ICU, Age, and BMI, respectively. For Scenario~2, as expected, the difference between the PTCR of \approach{} and the baseline grows as the patient’s criticality level, based on BMI, rises. This highlights the increasing importance of adaptation, which becomes even more pronounced as the patient’s condition becomes more critical. Scenario 3 intrinsically focuses on critical condition patients. Thus, as expected, the difference between the PTCR of \approach{} and the baseline is larger for patients considered at higher risk. Although all ICUs are important as they admit patients with varying degrees of criticality, we believe that the ICU for cardiac surgery patients is the most severe. Therefore, confirming this hypothesis, "CardiacSurgeryUnit" shows the greatest difference among all the ICU profiles. The same applies to "Age", as the "60+" patient profile is most impacted by the use of adaptation in this scenario. Finally, regarding the BMI, the greatest difference in PTCR was for "Obesity 1" and not "Obesity 3" as expected. However, this may be explained by the fact that it is not only the BMI that impacts the patient's condition but also other factors, such as age and lifestyle.

\subsection{Threats to Validity}
Despite our best efforts, potential threats to the validity of the results might remain. We present four validity key aspects following the classification in \cite{runeson2009guidelines} as follows. 
\\
\noindent {\em Construct validity}: whether the designed experiments are appropriate for assessing the effectiveness of \approach{} in all the scenarios.
Our validation demonstrates that \approach{} successfully identified previously unknown failures in SA-BSN and outperformed the Baseline in terms of effectiveness. While we used PTCR, a widely accepted metric for measuring effectiveness, we acknowledge that the use of alternative metrics could potentially yield different results. Besides, we compared \approach{} solely with our Baseline due to the scarce availability of comparable testing approaches. 
However, we consider the Baseline a strong choice in this case, as in this way we can demonstrate that incorporating adaptation logic into a testing approach can help uncover more failures. Another potential threat to validity not only in \approach{} but also in the Baseline approaches lies in utilizing DTMCs where the initial state is consistently set to the low-risk level. However, this choice follows the original implementation of SA-BSN.

\noindent {\em Internal validity}: whether the observed results were actually caused by the 
"treatment" rather than other factors. We adhered to established guidelines for conducting the experiments and performing statistical analysis. A common internal threat is the precision of measurements, which can be affected by random factors. To mitigate this potential issue, we repeated the experiments 5 times.

\noindent {\em External validity}: whether and to what extent the observations can be generalized. \approach{} is a broadly applicable approach, but for the sake of clarity, we focus on a BSN within the healthcare domain. Even so, to date, our results are solely based on the SA-BSN case study. 
Additional experiments are necessary to more thoroughly validate the effectiveness of \approach{} and explore its potential application in other BSN domains, or even, other evolving systems.

\noindent {\em Reliability}: Whether and to what degree other researchers can reproduce the results. To ensure reproducibility, we provide all data and configuration details in a replication package. 

\section{Related Work}\label{sec:relatedwork}

\approach{} is proposed as a model-based approach for Self-Adaptive Testing in the Field (SATF) of BSNs. 
Indeed, SATF is a branch of testing that is recently drawing interest, yet so far has not been extensively explored in the literature. For comprehensiveness, we can leverage the already cited systematic review in~\cite{silva2024self}, which 
collected 19 papers over the period from 2012 until mid-2022 (due to lack of space, we refer to the cited review for their complete list). In search for more SATF papers that could have appeared after 2022, 
we launched their same query on the same repositories, filtering for the period 2022-2024. By applying the same selection criteria used in~\cite{silva2024self}, we eventually found two more papers (namely \cite{haq2023many} and~\cite{adigun2023risk}) that added to the original 19 ones make a total of 21 papers proposing SATF approaches.




We then reviewed these 21 papers: first, we noticed that while many of them are employed to test adaptive systems, 
{it would require considerable effort to adapt them for application to BSNs and for comparison against \approach{}}. Then, 
{we considered the most similar works more specifically looking at the features exposed,}
including: the continuous use of a monitor to detect the possible need for adaptation; the type of adaptations that are performed; no human in the loop to manage the adaptation; possibly application of a model-based approach for test generation. 
For instance,  Gazzola et al.~\cite{gazzola2022testing} use data collected from the field to generate test cases that could trigger yet unexplored execution states. In principle, their intent is similar to what we do by using patient data to build the DTMC that is then used to test the BSN. However, their approach targets unit testing and is conceived for testing in production, while we conduct ex-vivo testing at system level.

SAMBA~\cite{leal2019samba} is a Self-Adaptive Model-BAsed online testing framework addressing Service Oriented Architecture (SOA). When its monitor detects a change in the SOA under test, the Model Generator component is responsible for modifying the model that is referred for test generation. Then the new generated test cases are executed. The overall scheme is partly similar to our approach functioning for what concerns our adaptation of test cases and/or test oracle. However, the types of system targeted (SOA vs. BSN) differ widely, and consequently, the implementation of test generation and execution modules are completely different. 
Moreover, they execute test cases in production and not ex-vivo.

Perhaps the most similar approach is the framework  WS-REPAS~\cite{guerriero2019hybrid}, which combines a DTMC modeling approach with monitoring for in vivo testing of services. 
Precisely, they collect field data through monitoring and use them to continuously update the parameters
of a DTMC model. When changes are observed, testing is triggered, which executes an in-vivo testing session.
Differently from \approach{}, though,  WS-REPAS aims at non-functional testing.

Concerning approaches that are specifically conceived for testing BSNs, we found in the literature works conducting fault diagnosis to gather confidence in BSNs \cite{zhang2017fault,Mahapatro2012FaultDI}. The work in~\cite{zhang2017fault}
proposes a method that utilizes Hidden Markov Models (HMMs) to detect faulty readings from individual sensors. Similarly, the study in~\cite{Mahapatro2012FaultDI} uses multi-sensor fusion approaches to identify faulty sensors by assuming correlations between sensorial measurements and diseases. However, neither of them is performing testing.
Finally, our previous work in~\cite{silva4876283different} outlines a strategy for testing BSNs. Indeed, \approach{}, previously described in Figure \ref{fig:method}, is partly inspired by their GATE4BSN approach. However, GATE4BSN lacks field data and does not adapt the testing strategy as \approach{} does.

\section{Discussion and Conclusions}\label{sec:conc}

This paper introduces \approach{}, a new approach for testing BSN applications. \approach{} employs an ex-vivo approach, using real-world data collected from the field to
simulate patient behavior in in-house experiments. We derive DTMCs to simulate different patient profiles and use them to generate the test input data for the BSN. The outputs of the BSN are compared with a proposed oracle to assess the test results. \approach{} adaptive logic continuously observes the SUT and the simulated patient, making testing adaptations when needed. We also presented three adaptation scenarios, detailing the trigger conditions and the adaptation policies for each.

We assessed \approach{} 
using a self-adaptive BSN system and a clinical database from the literature and identified several previously unknown failures. Through experiments, we compared \approach{} against a baseline, which is a non-adaptive version of the proposed approach for each of the scenarios. The results, confirmed by statistical analysis, demonstrated that AdapTA achieves greater effectiveness compared to the baseline across all three adaptation scenarios, emphasizing the value of its adaptive logic. Notably, scenario 3, which focused the testing on the patient's critical situation, showed the best improvement. This scenario likely represents the context where the SA-BSN should perform optimally, as missing an alarm could have serious consequences, hence it is crucial to give this scenario appropriate attention.
Through the analysis of the rate of passing  tests by patient profile,  
we noticed that for Scenarios 2  and 3, which take into consideration the patient criticality, the adaptation benefited the most the most critical patient profiles. On the other hand, for Scenario 1, which does not take into consideration the patient's conditions to perform the adaptation, we could not see a clear impact of the patient profile on the improvement brought by the adaptation. 

In future work, we plan to assess \approach{} on other BSNs in different domains. The proposed method could also be used to test evolving systems like Cyber-Physical Systems and autonomous systems, which depend on specific input sequences that can lead to failures. Since the adaptation logic is external to the testing components, \approach{} can be easily modified to incorporate new scenarios, expand current ones to include different patient profiles, provide more accurate oracles (e.g., from domain experts), and explore alternative testing strategies.

\section*{Acknowledgment}
This work has been (partially) funded by (i) the MUR (Italy) -- PRIN PNRR 2022 project ``RoboChor: Robot Choreography'' (grant P2022RSW5W), (ii) the European Union - NextGenerationEU under the Italian Ministry of University and Research (MUR) National Innovation Ecosystem grant ECS00000041 - VITALITY – CUP: D13C21000430001, (iii) The MUR (PNRR) and ASI ``Space it up" project, and (iv) the European HORIZON-KDT-JU-2023-2-RIA research project MATISSE ``Model-based engineering of Digital Twins for early verification and validation of Industrial Systems" (grant  101140216-2, KDT232RIA\_00017). This work is also supported by the Wallenberg AI, Autonomous Systems and Software Program (WASP) funded by the Knut and Alice Wallenberg Foundation.

\bibliographystyle{IEEEtran}
\bibliography{references}
\end{document}